\documentclass[11pt]{article}
\usepackage{amsmath}
\usepackage{epsfig}

\newcommand{\ord}[1]{\mathcal{O}(#1)}
\newcommand{\del}{\partial}

\newcommand{\e}[1]{\text{exp}_\star [ \, #1 \, ]}
\newcommand{\ket}[1]{|#1\rangle}
\newcommand{\dcirc}{\mathop{{}^\circ_\circ}}

\newcommand{\state}{\Omega}

\begin{document}

\baselineskip=17pt

\begin{titlepage}
\rightline{\tt arXiv:0708.3394}
\rightline{\tt DESY 07-130}
\rightline{\tt MIT-CTP-3857}
\begin{center}

\vskip 2.0cm
{\Large \bf {General marginal deformations in open
superstring field theory}}\\
\vskip 1.5cm

{\large {Michael Kiermaier${}^1$ and Yuji Okawa${}^2$}}

\vskip 1.0cm

{\it {${}^1$ Center for Theoretical Physics}}\\
{\it {Massachusetts Institute of Technology}}\\
{\it {Cambridge, MA 02139, USA}}\\
mkiermai@mit.edu

\vskip 0.5cm

{\it {${}^2$ DESY Theory Group}}\\
{\it {Notkestrasse 85}}\\
{\it {22607 Hamburg, Germany}}\\
yuji.okawa@desy.de

\vskip 1.5cm

{\bf Abstract}
\end{center}

\noindent
We construct analytic solutions
of open superstring field theory
for any exactly marginal deformation
in any boundary superconformal field theory
when properly renormalized operator products
of the marginal operator are given.
Our construction is an extension of the general framework
for marginal deformations developed in arXiv:0707.4472
for open bosonic string field theory,
and the solutions are based on integrated vertex operators
which are closely related to
finite deformations in boundary superconformal field theory.

\end{titlepage}

\newpage

\tableofcontents

\section{Introduction}
\setcounter{equation}{0}

The purpose of the paper is to extend the general framework
for marginal deformations
developed in~\cite{Kiermaier:2007vu}
for open bosonic string field theory~\cite{Witten:1985cc}
to open superstring field theory formulated
by Berkovits~\cite{Berkovits:1995ab}.\footnote{
See \cite{Taylor:2003gn, Sen:2004nf, Rastelli:2005mz, Taylor:2006ye}
for reviews on string field theory.}
Let us briefly review recent remarkable progress
in analytic methods for open string field
theory~\cite{Schnabl:2005gv}--\cite{Rastelli:2007gg},
focusing on marginal deformations.
Analytic solutions for marginal deformations
were first constructed in~\cite{Schnabl:2007az, Kiermaier:2007ba}
for the bosonic string
when operator products of the marginal operator are regular,
and the solutions were extended to the superstring
in~\cite{Erler:2007rh, Okawa:2007ri, Okawa:2007it}.
The generalization to marginal deformations
with singular operator products was initiated
in~\cite{Kiermaier:2007ba}, and solutions to third order
in the deformation parameter were constructed.
For the special case of the marginal deformation
corresponding to the zero mode of the gauge field,
solutions to all orders were constructed
for the bosonic string in~\cite{Fuchs:2007yy}
and for the superstring in~\cite{Fuchs:2007gw}.
The solutions in~\cite{Fuchs:2007yy,Fuchs:2007gw}, however,
do not satisfy the reality condition on the string field,
and a strategy for constructing real solutions
was outlined in~\cite{Fuchs:2007gw}.
See~\cite{Sen:2000hx}--\cite{Jokela:2007dq}
for earlier study of marginal deformations
in string field theory and related work.

Analytic solutions for general marginal deformations
satisfying the reality condition were recently
constructed in~\cite{Kiermaier:2007vu}
for the bosonic string.
While previous solutions for marginal deformations
in~\cite{Schnabl:2007az, Kiermaier:2007ba,
Erler:2007rh, Okawa:2007ri, Okawa:2007it}
were built from
unintegrated vertex operators
and $b$-ghost insertions,
the solutions in~\cite{Kiermaier:2007vu} were
based on integrated vertex operators
which are closely related to finite deformations
of boundary conformal field theory (CFT).
A change of boundary conditions
in boundary CFT
can be implemented
by properly renormalized exponential operators
of an integral of the marginal operator,
and a systematic procedure to construct solutions from
such renormalized operators was presented
in~\cite{Kiermaier:2007vu}.
The general idea of the construction
in~\cite{Kiermaier:2007vu} does not depend
on the bosonic nature of the problem,
and we expect that the construction can be
extended to the superstring.
We in fact find that the extension is remarkably simple,
and we construct analytic solutions
of open superstring field theory
to all orders in the deformation parameter
satisfying the reality condition.

The organization of the paper is as follows.
In section~\ref{sec2} we review the construction
in~\cite{Kiermaier:2007vu}
of solutions to the equation of motion
for the bosonic string.
We use this result later
and construct string fields in the superstring
satisfying the bosonic equation of motion
with the BRST operator in the bosonic theory
replaced by the one in the superstring theory.
In section~\ref{sec3} we discuss properties
of integrated vertex operators in the superstring.
In section~\ref{sec4} we construct
solutions to the equation of motion
of open superstring field theory.
This is the main result of the paper.
String field theory expanded around the solution
was described in~\cite{Kiermaier:2007vu}
using a deformed star product.
In section~\ref{sec5} we show that
the equation of motion of open superstring field theory
expanded around the solution in section~\ref{sec4}
can also be described using the deformed star product
in~\cite{Kiermaier:2007vu}.
Section~\ref{sec6} is devoted to discussion.

\section{Solutions to the bosonic equation of motion}\label{sec2}
\setcounter{equation}{0}

The equation of motion of open bosonic string field
theory~\cite{Witten:1985cc} is given by
\begin{equation}\label{boseom}
Q_B \Psi + \Psi^2 = 0 \,,
\end{equation}
where $\Psi$ is the string field of ghost number one
and $Q_B$ is the BRST operator.
Here and in what follows products of string fields are defined
by the star product~\cite{Witten:1985cc}.
In this section we review the construction
in~\cite{Kiermaier:2007vu}
of solutions to~(\ref{boseom})
for general marginal deformations.

A marginal deformation is generated by a marginal operator $V_1(t)$
which is a matter primary field of dimension one.
The solutions in~\cite{Kiermaier:2007vu} are constructed from
an operator which implements a change of boundary conditions
between two points $a$ and $b$ on the boundary.
When operator products of the marginal operator are regular,
it is given by
\begin{equation}\label{exp-before-renormalization}
\exp \biggl[ \, \lambda \int_a^b dt \, V_1(t) \, \biggr]
= 1 + \lambda \int_a^b dt \, V_1(t)
+ \frac{\lambda^2}{2!} \int_a^b dt_1 \int_a^b dt_2 \,
V_1(t_1) \, V_1(t_2) + \, \ldots \,,
\end{equation}
where $\lambda$ is the deformation parameter.
When operator products of the marginal operator are singular,
we need to renormalize
the operator~(\ref{exp-before-renormalization})
properly to make it well defined,
and we denote the renormalized operator by
\begin{equation}\label{[exp]_r}
[ \, e^{\lambda V(a,b)} \, ]_r \,,
\end{equation}
where
\begin{equation}
V(a,b) \equiv \int_a^b dt \, V_1(t) \,.
\end{equation}
If the marginal deformation is exactly marginal,
there is a one-parameter family of consistent boundary conditions
labeled by $\lambda$
and we expect to have a corresponding
family of solutions
in string field theory.
Since the new boundary condition
generated by the operator $[ \, e^{\lambda V(a,b)} \, ]_r$
is conformal,
the operator $[ \, e^{\lambda V(a,b)} \, ]_r$ should be
invariant under the BRST transformation
up to additional contributions from
the points $a$ and $b$ where the boundary condition changes:
\begin{equation}\label{I}
Q_B \cdot [ \, e^{\lambda V(a,b)} \, ]_r
= [ \, e^{\lambda V(a,b)} \, O_R (b) \, ]_r
- [ \, O_L (a) \, e^{\lambda V(a,b)} \, ]_r \,.
\end{equation}
Here $O_L (a)$ and $O_R (b)$ are some local operators
at $a$ and $b$, respectively.
See the introduction of~\cite{Kiermaier:2007vu}
for more detailed discussion.
The solutions in~\cite{Kiermaier:2007vu}
were constructed from the operator
$[ \, e^{\lambda V(a,b)} \, ]_r$ as follows.
The operator $[ \, e^{\lambda V(a,b)} \, ]_r$ is given
in the form of an expansion in $\lambda$:
\begin{equation}
[ \, e^{\lambda V(a,b)} \, ]_r
= \sum_{n=0}^\infty \lambda^n \,
[ \, V^{(n)}(a,b)\, ]_r \,,
\end{equation}
where
\begin{equation}
[ \, V^{(n)} (a,b)\, ]_r
\equiv \frac{1}{n!} \, \bigl[ \, \bigl( V(a,b) \bigr)^n \, \bigr]_r
\quad\text{ for }\quad n\geq 1
\quad\text{ and }\quad [\,V^{(0)}(a,b)\,]_r\equiv1 \,.
\end{equation}
We then define a state $U$ by
\begin{equation}
U \equiv 1 + \sum_{n=1}^\infty \, \lambda^n \, U^{(n)} \,,
\end{equation}
where
\begin{equation}
\langle \, \varphi \,,\, U^{(n)} \, \rangle
= \langle \, f \circ \varphi (0) \, \,
[ \, V^{(n)} (1,n) \, ]_r \, \rangle_{{\cal W}_n} \,.
\end{equation}
Here and in what follows we denote a generic state
in the Fock space by $\varphi$ and its corresponding operator
in the state-operator mapping by $\varphi (0)$.
The conformal transformation $f(\xi)$ is
\begin{equation}
f(\xi) = \frac{2}{\pi} \, \arctan \xi \,,
\end{equation}
and we denote the conformal transformation of
the operator $\varphi (\xi)$
under the map $f(\xi)$ by $f \circ \varphi (\xi)$.
The correlation function is evaluated
on the surface ${\cal W}_n$
which is obtained from the upper-half plane of $z$
by the identification $z \sim z + n + 1$.
We represent ${\cal W}_n$ in the region
where $-1/2 \le {\rm Re} \, z \le 1/2 + n$.
It follows from~(\ref{I}) that
the BRST transformation of the operator
$[ \, V^{(n)}(a,b) \, ]_r$ takes the form
\begin{equation}
Q_B \cdot [ \, V^{(n)}(a,b) \, ]_r
= \sum_{r=1}^n \, [  \, V^{(n-r)}(a,b) \, O_R^{(r)}(b)\, ]_r
{}- \sum_{l=1}^n \, [ \, O_L^{(l)}(a) \, V^{(n-l)}(a,b) \, ]_r\,,
\end{equation}
where $O_L$ and $O_R$ are expanded as follows:
\begin{equation}
O_L = \sum_{n=1}^\infty \, \lambda^n \, O_L^{(n)} \,, \qquad
O_R = \sum_{n=1}^\infty \, \lambda^n \, O_R^{(n)} \,.
\end{equation}
Thus the BRST transformation of $U$ can be split into two pieces:
\begin{equation}\label{QU}
Q_B U = A_R - A_L
\end{equation}
with
\begin{equation}
A_L = \sum_{n=1}^\infty \, \lambda^n \, A_L^{(n)} \,, \qquad
A_R = \sum_{n=1}^\infty \, \lambda^n \, A_R^{(n)} \,,
\end{equation}
where
\begin{equation}\label{A_L/R}
\begin{split}
    \langle \, \varphi \,, A_L^{(n)} \, \rangle &=
    \sum_{l=1}^n \langle \, f \circ \varphi (0) \,
    [ \, O_L^{(l)} (1) \, V^{(n-l)} (1,n) \, ]_r \,
    \rangle_{{\cal W}_n} \,, \\
    \langle \, \varphi \,, A_R^{(n)} \, \rangle &=
    \sum_{r=1}^n \langle \, f \circ \varphi (0) \,
    [  \, V^{(n-r)} (1,n) \, O_R^{(r)} (n)\, ]_r \,
    \rangle_{{\cal W}_n}  \,.
\end{split}
\end{equation}
We then define $\Psi_L$ by
\begin{equation}
\Psi_L \equiv A_L \, U^{-1} \,,
\end{equation}
where $U^{-1}$ is well defined perturbatively in $\lambda$
because $U=1+\ord{\lambda}$.
The BRST transformation of $\Psi_L$ can be calculated as follows:
\begin{equation}
\begin{split}
    Q_B\Psi_L \, &= \, Q_B\,(A_L \, U^{-1})\\
    &=(Q_B A_L)\, U^{-1}+A_L\, U^{-1}\,(Q_B U)\, U^{-1}\\
    &=(Q_B A_L)\, U^{-1}+A_L\, U^{-1}\,(A_R-A_L)\, U^{-1}\\
    &=(Q_B A_L +A_L\, U^{-1}\, A_R)\, U^{-1}-A_L\, U^{-1}\, A_L\, U^{-1}\\
    &=(Q_B A_L +A_L\, U^{-1}\, A_R)\, U^{-1}-\Psi_L^2 \,.
\end{split}
\end{equation}
It was shown in~\cite{Kiermaier:2007vu} that the relation
\begin{equation}\label{Q_B-A_L}
Q_B A_L = {}- A_L\, U^{-1}\, A_R
\end{equation}
holds under a set of assumptions
which were argued to be satisfied
for any exactly marginal deformation.
The equation~(\ref{I}) is in fact the first of these assumptions.
We list the complete
set of assumptions in appendix~\ref{assumptions}.
The state $\Psi_L$ thus solves the equation of motion:
\begin{equation}
Q_B \Psi_L + \Psi_L^2 = 0 \,.
\end{equation}

The solution $\Psi_L$, however, does not satisfy
the reality condition on the string field,
and a solution satisfying the reality condition was
generated in~\cite{Kiermaier:2007vu} from $\Psi_L$
by a gauge transformation.
The string field $\Psi$
must have a definite parity
under the combination of the Hermitean conjugation (hc)
and the inverse BPZ conjugation ($\text{bpz}^{-1}$)
to guarantee that the string field theory action
is real~\cite{Gaberdiel:1997ia}.
We define the conjugate $X^\ddagger$ of a string field $X$ by
\begin{equation}
X^\ddagger \equiv \text{bpz}^{-1} \circ \text{hc} \, (X) \,.
\end{equation}
The conjugation satisfies
\begin{eqnarray}
(Q_B X)^\ddagger &=& {}- (-1)^X \, Q_B X^\ddagger \,,
\label{conjugation-with-Q_B} \\
(X \, Y)^\ddagger &=& Y^\ddagger \, X^\ddagger \,.
\end{eqnarray}
Here and in what follows a string field in the exponent of $(-1)$
denotes its Grassmann property:
it is $0$ mod $2$ for a Grassmann-even state
and $1$ mod $2$ for a Grassmann-odd state.
In order for $Q_B \Psi$ and $\Psi^2$ to have the same
conjugation property, the Grassmann-odd string field $\Psi$
must satisfy $\Psi^\ddagger = \Psi$.
This is the reality condition on the string field
in open bosonic string field theory.
When the renormalized operator
$[ \, e^{\lambda V(a,b)} \, ]_r$
preserves the invariance under the reflection
where $V_1(t)$ is replaced by $V_1(a+b-t)$
and when $V_1$ is chosen such that
the state corresponding to $\lambda\,V_1(0)$
is even under the conjugation,\footnote{
If the state corresponding to $V_1(0)$
is odd under the conjugation,
we set $\lambda = i \, \tilde{\lambda}$
and take $\tilde{\lambda}$ to be real
to satisfy this convention.}
we have
\begin{equation}\label{someconjugations}
U^\ddagger = U \,, \qquad (U^{-1})^\ddagger = U^{-1} \,, \qquad
A_L^\ddagger = A_R \,.
\end{equation}
Therefore, a state $\Psi_R$ defined by
\begin{equation}
\Psi_R \equiv U^{-1} A_R
\end{equation}
is the conjugate of $\Psi_L$ and solves the equation of motion.
The two solutions $\Psi_L$ and $\Psi_R$ are related
by the gauge transformation generated by $U$:
\begin{equation}
\Psi_R \, = \, U^{-1} \, \Psi_L \, U + U^{-1} \, Q_B U \,.
\label{Psi_L-to-Psi_R}
\end{equation}
A solution $\Psi$ satisfying the reality condition
is obtained from $\Psi_L$ or $\Psi_R$
by gauge transformations as follows:
\begin{equation}\label{Psireal}
\begin{split}
\Psi & = \frac{1}{\sqrt{U}} \, \Psi_L \, \sqrt{U} +
\frac{1}{\sqrt{U}} \, Q_B \sqrt{U} \\
&= \sqrt{U} \, \Psi_R \, \frac{1}{\sqrt{U}}
+ \sqrt{U} \, Q_B \frac{1}{\sqrt{U}} \\
& = \frac{1}{2} \, \biggl[ \, \frac{1}{\sqrt{U}} \, \Psi_L \,
\sqrt{U} + \sqrt{U} \, \Psi_R \, \frac{1}{\sqrt{U}} +
\frac{1}{\sqrt{U}} \, Q_B  \sqrt{U}
- (Q_B  \sqrt{U}) \, \frac{1}{\sqrt{U}} \, \biggr] \,,
\end{split}
\end{equation}
where $\sqrt{U}$ and $1/\sqrt{U}$ are defined perturbatively
in $\lambda$.
It follows from $( \sqrt{U} \, )^\ddagger = \sqrt{U}$,
$( 1/\sqrt{U} \, )^\ddagger = 1/\sqrt{U}$
and $\Psi_L^\ddagger=\Psi_R$ that
the last expression for $\Psi$ in~(\ref{Psireal}) manifestly
satisfies the reality condition.
The three expressions are equivalent
because of the relation (\ref{Psi_L-to-Psi_R}).

\section{Integrated vertex operators in the superstring}\label{sec3}
\setcounter{equation}{0}

We expect that integrated vertex operators play
a crucial role
in extending the construction of solutions
in~\cite{Kiermaier:2007vu} to the superstring.
The marginal operator $V_1$ in the superstring is
the supersymmetry transformation
of a superconformal primary field $\widehat{V}_{1/2}$
in the matter sector
of dimension $1/2$:
\begin{equation}\label{ssV1}
V_1(t) = G_{-1/2} \cdot \widehat{V}_{1/2}(t) \equiv
\int_{C(t)} \Bigl[ \, \frac{dz}{2\pi i} \, T_F(z) \,
- \frac{d \bar{z}}{2 \pi i} \, \widetilde{T}_F(\bar{z}) \,
\Bigr] \, \widehat{V}_{1/2}(t) \,,
\end{equation}
where $T_F(z)$ and $\widetilde{T}_F(\bar{z})$ are
the holomorphic and antiholomorphic components, respectively,
of the world-sheet supercurrent,
and $C(t)$ is a contour in the upper-half plane
which runs from the point $t+\epsilon$
on the real axis to the point $t-\epsilon$ on the real axis
in the limit $\epsilon \to 0$ with $\epsilon > 0$.
An integrated vertex operator
in the $0$ picture
is an integral of $V_1$ on the boundary:
\begin{equation}
V(a,b) =\int_a^b dt\, V_1(t)
= \int_a^b dt \, G_{-1/2} \cdot \widehat{V}_{1/2}(t) \,.
\end{equation}
It is invariant under the BRST transformation
up to nonvanishing terms from the end points
of the integral region:
\begin{equation}\label{QVab}
\begin{split}
Q_B \cdot V(a,b)
&= \int_a^b dt \,\del_t \,
[ \, cV_1(t) + \eta e^\phi \, \widehat{V}_{1/2}(t) \, ]\\
&= [ \, cV_1(b) + \eta e^\phi \, \widehat{V}_{1/2}(b) \, ]
- [ \, cV_1(a) + \eta e^\phi \, \widehat{V}_{1/2}(a) \, ] \,.
\end{split}
\end{equation}
We use the description of the superconformal ghosts
in terms of $\eta$, $\xi$,
and $\phi$~\cite{Friedan:1985ge, Polchinski:1998rr},
and the BRST operator for the superstring is given by
\begin{equation}\label{Q_B}
Q_B = \int \Bigl[ \, \frac{dz}{2 \pi i} \, j_B (z)
- \frac{d \bar{z}}{2 \pi i} \, \tilde{\jmath}_B (\bar{z}) \, \Bigr]
\end{equation}
with
\begin{equation}
\begin{split}
j_B = c \, T_B^m + c \, T_B^{\eta \xi} + c \, T_B^\phi
+ \eta e^\phi \, T_F^m
+ {}: b c \partial c :
{}- b \eta \partial \eta e^{2 \phi} \,, \\
T_B^{\eta \xi} = - \eta \partial \xi \,, \qquad
T_B^\phi = -\frac{1}{2} \, \partial \phi \partial \phi
- \partial^2 \phi \,,
\end{split}
\end{equation}
where $T_B^m$ and $T_F^m$ are the holomorphic components
of the energy-momentum tensor and the supercurrent
in the matter sector, respectively,
and $\tilde{\jmath}_B$ is
the antiholomorphic counterpart of $j_B$.
The operator $V(a,b)$ in the matter sector is obviously
annihilated by $\eta_0$,
which is the zero mode of $\eta$
and plays an important role in open superstring field
theory~\cite{Berkovits:1995ab}.
Since the BRST operator anticommutes with $\eta_0$,
the operator $Q_B \cdot V(a,b)$ is also annihilated by $\eta_0$.
We can explicitly see that
the operator $cV_1(t) + \eta e^\phi \, \widehat{V}_{1/2}(t)$
which appeared in~(\ref{QVab}) is annihilated by $\eta_0$.
The operator $cV_1(t) + \eta e^\phi \, \widehat{V}_{1/2}(t)$
is also annihilated by the BRST operator.
This can be seen by acting with $Q_B$ on~(\ref{QVab}).
In the description of the superconformal ghosts
in terms of $\eta$, $\xi$, and $\phi$
including the sector generated by $\eta_0$ and $\xi_0$,
any BRST-closed operator
can be written as an BRST-exact operator
because of the existence of
a Grassmann-odd operator
$R(t)$ satisfying
\begin{equation}\label{QBghosts=1}
Q_B \cdot R(t) = 1 \,.
\end{equation}
See, for example, footnote 3 of~\cite{Berkovits:2004xh}. We choose $R(t)$ to be
\begin{equation}
R(t) \equiv {}- c \xi \partial \xi e^{-2 \phi}(t) \,.
\end{equation}
Since
\begin{equation}\label{unintegrated-vertex-operator}
\lim_{\epsilon \to 0} R(t-\epsilon) \,
[ \, cV_1(t) + \eta e^\phi \, \widehat{V}_{1/2}(t) \, ]
= c \xi e^{-\phi}\widehat{V}_{1/2}(t) \,,
\end{equation}
we have
\begin{equation}\label{O1=QO1hat}
cV_1(t) + \eta e^\phi \, \widehat{V}_{1/2}(t)
= Q_B \cdot \bigl[ \,
c \xi e^{-\phi}\widehat{V}_{1/2}(t) \, \bigr] \,.
\end{equation}
Note that the unintegrated vertex operator
$c e^{-\phi}\widehat{V}_{1/2}$ in the $-1$ picture
with an additional factor of $\xi$
appeared in~(\ref{unintegrated-vertex-operator}).
This operator is used in the solution
to the linearized equation of motion
of open superstring field theory
formulated by Berkovits~\cite{Berkovits:1995ab},
as we will discuss in the next section.

Finite deformations of the boundary CFT are generated
by an exponential of $V(a,b)$.
When operator products of $V_1$ are regular, it is given by
$e^{\lambda V(a,b)}$, where $\lambda$ is the deformation parameter.
Its BRST transformation is
\begin{equation}\label{regQexpVab}
Q_B \cdot e^{\lambda V(a,b)}
= \lambda \, e^{\lambda V(a,b)} \,
[ \, cV_1(b) + \eta e^\phi \, \widehat{V}_{1/2}(b) \, ]
- \lambda \, [ \, cV_1(a) + \eta e^\phi \, \widehat{V}_{1/2}(a) \, ] \,
e^{\lambda V(a,b)}
\end{equation}
if operator products of $\widehat{V}_{1/2}$
and an arbitrary number of $V_1$'s are also regular
so that~(\ref{QVab}) can be applied even in the presence
of further insertions of $V_1$'s.
The second term on the right-hand side
can be written as
\begin{equation}
\begin{split}
& \lambda \, [ \, cV_1(a) + \eta e^\phi \, \widehat{V}_{1/2}(a) \, ] \,
e^{\lambda V(a,b)}
\\&
= \lambda\, Q_B \cdot
\bigl[ \, c\xi e^{-\phi}\widehat{V}_{1/2}(a) \, \bigr] \,
e^{\lambda V(a,b)} \\
& = \lambda \, Q_B \cdot \bigl[ \, c\xi e^{-\phi}\widehat{V}_{1/2}(a) \,
e^{\lambda V(a,b)} \, \bigr] \,
- \lambda^2 \,
[ \, c\xi e^{-\phi}\widehat{V}_{1/2}(a) \,] \, e^{\lambda V(a,b)} \,
[ \, cV_1(b) + \eta e^\phi \, \widehat{V}_{1/2}(b) \, ] \,,
\end{split}
\end{equation}
where we again used the regularity assumption
on the operator products.
We thus find that
\begin{equation}\label{assVIIregular}
\begin{split}
& \lambda \, Q_B \cdot \bigl[ \,
c\xi e^{-\phi}\widehat{V}_{1/2}(a) \, e^{\lambda V(a,b)} \, \bigr] \\
& = \lambda \,
[ \, cV_1(a) + \eta e^\phi \, \widehat{V}_{1/2}(a) \, ] \,
e^{\lambda V(a,b)}
+ \lambda^2 \, [ \, c\xi e^{-\phi}\widehat{V}_{1/2}(a) \, ] \,
e^{\lambda V(a,b)} \,
[ \, cV_1(b) + \eta e^\phi \, \widehat{V}_{1/2}(b) \, ] \,.
\end{split}
\end{equation}
This relation, being generalized to the singular case, plays
a crucial role in our construction
of solutions in open superstring field theory.

When operator products of $V_1$ are singular,
we need to renormalize the operator $e^{\lambda V(a,b)}$
properly to make it well defined,
and we denote the renormalized operator
by $[ \, e^{\lambda V(a,b)} \, ]_r$\
as before.
If the deformation is exactly marginal and preserves
superconformal invariance,
we assume as in the bosonic case that
the BRST transformation of $[ \, e^{\lambda V(a,b)} \, ]_r$
takes the following form:
\begin{equation}
Q_B \cdot [ \, e^{\lambda V(a,b)} \, ]_r
= [ \, e^{\lambda V(a,b)} \, O_R(b) \, ]_r
- [ \, O_L(a) \, e^{\lambda V(a,b)} \, ]_r \,,
\end{equation}
where $O_L(a)$ and $O_R(b)$ are some Grassmann-odd local operators
at $a$ and $b$, respectively.
The operators $[ \, O_L(a) \, e^{\lambda V(a,b)} \, ]_r$
and $[ \, e^{\lambda V(a,b)} \, O_R(b) \, ]_r$
are annihilated by $\eta_0$\,, as we discussed before.
To leading order in $\lambda$
they are determined from~(\ref{QVab}) and given by
\begin{equation}\label{leadingOs}
\begin{split}
[ \, O_L(a) \, e^{\lambda V(a,b)} \, ]_r
& = \lambda \, [ \, cV_1(a) + \eta e^\phi \, \widehat{V}_{1/2}(a) \, ]
+ \ord{\lambda^2} \,, \\
[ \, e^{\lambda V(a,b)} \, O_R(b) \, ]_r
& = \lambda \, [ \, cV_1(b) + \eta e^\phi \, \widehat{V}_{1/2}(b) \, ]
+ \ord{\lambda^2} \,.
\end{split}
\end{equation}
In the regular case,
we find from the exact expression in~(\ref{regQexpVab}) that
\begin{equation}
O_L^{regular} = O_R^{regular} =
\lambda \, [ \, cV_1 + \eta e^\phi \, \widehat{V}_{1/2} \, ] \,,
\end{equation}
and there are no higher-order corrections to the operators
$O_L$ and $O_R$.

Let us introduce the following operators:
\begin{equation}\label{defOhatExp}
\begin{split}
[ \, \widehat{O}_L(a) \, e^{\lambda V(a,b)} \, ]_r&\equiv \quad
\lim_{\epsilon\to 0}\, R (a-\epsilon)\,
[ \, O_L(a) \, e^{\lambda V(a,b)} \, ]_r\,,\\
[ \,  e^{\lambda V(a,b)}\,\widehat{O}_R(b) \, ]_r&\equiv
-\lim_{\epsilon\to 0}\,[  \, e^{\lambda V(a,b)} \, O_R(b) \, ]_r\,
R (b+\epsilon)\,.
\end{split}
\end{equation}
These are generalizations of
$\lambda \, c\xi e^{-\phi}\widehat{V}_{1/2}(a) \, e^{\lambda V(a,b)}$
and $\lambda \, e^{\lambda V(a,b)} \, c\xi e^{-\phi}\widehat{V}_{1/2}(b)$
in the regular case.
The ghost sector couples to the matter sector
only through $c$ and $\eta e^{\phi}$ in the BRST current,
and the operator products of $c \xi \partial \xi e^{-2 \phi}$
with $c$, $\eta e^{\phi}$, and their derivatives are regular.
The limit $\epsilon \to 0$
in~(\ref{defOhatExp}) is therefore regular.
To leading order in $\lambda$ these operators reduce to
\begin{equation}\label{hatO-leading}
\begin{split}
[ \, \widehat{O}_L(a) \, e^{\lambda V(a,b)} \, ]_r
& = \lambda \, c \xi e^{-\phi} \widehat{V}_{1/2}(a) + \ord{\lambda^2} \,, \\
[ \,  e^{\lambda V(a,b)}\,\widehat{O}_R(b) \, ]_r
& = \lambda \, c \xi e^{-\phi} \widehat{V}_{1/2}(b) + \ord{\lambda^2} \,.
\end{split}
\end{equation}
The BRST transformation of
$[ \, \widehat{O}_L(a) \, e^{\lambda V(a,b)} \, ]_r$
can be calculated from~(\ref{QBghosts=1}) and from
the BRST transformation of
$[ \, O_L(a) \, e^{\lambda V(a,b)} \, ]_r$.
When the deformation is exactly marginal
and preserves superconformal invariance,
we assume that the BRST transformation of
$[ \, O_L(a) \, e^{\lambda V(a,b)} \, ]_r$
is given by
\begin{equation}
Q_B \cdot [ \, O_L (a) \, e^{\lambda V(a,b)} \, ]_r
= {}- [ \, O_L (a) \, e^{\lambda V(a,b)} \, O_R (b) \, ]_r \,.
\end{equation}
See the introduction of ~\cite{Kiermaier:2007vu}
for more detailed discussion.
The BRST transformation of
$[ \, \widehat{O}_L(a) \, e^{\lambda V(a,b)} \, ]_r$
is then given by
\begin{equation}\label{noassump1}
\begin{split}
Q_B\cdot[ \, \widehat{O}_L(a) \, e^{\lambda V(a,b)} \, ]_r
& = Q_B\cdot \Bigl[ \,
\lim_{\epsilon\to 0} R (a-\epsilon)\,
[ \, O_L(a) \, e^{\lambda V(a,b)} \, ]_r \, \Bigr] \\
&=[ \, O_L(a) \, e^{\lambda V(a,b)} \, ]_r
+\lim_{\epsilon\to 0} R (a-\epsilon)\,
[ \, O_L(a) \, e^{\lambda V(a,b)} \, O_R(b) \, ]_r \,.
\end{split}
\end{equation}
This is a generalization of~(\ref{assVIIregular}).
Similarly, we find
\begin{equation}\label{noassump2}
\begin{split}
Q_B\cdot[ \, e^{\lambda V(a,b)}\, \widehat{O}_R(b)  \, ]_r
& ={}- Q_B\cdot \Bigl[ \, \lim_{\epsilon \to 0} \,
[ \, e^{\lambda V(a,b)}\, O_R(b)  \, ]_r \,
R (b+\epsilon) \, \Bigr] \\
&=[  \, e^{\lambda V(a,b)} \, O_R(b)\, ]_r
+ \lim_{\epsilon \to 0} \,
[ \, O_L(a) \, e^{\lambda V(a,b)} \,O_R(b)\, ]_r \,
R (b+\epsilon) \,.
\end{split}
\end{equation}
As we mentioned before,
the relations~(\ref{noassump1}) and~(\ref{noassump2})
play a crucial role in the construction of the general
superstring solutions in the next section.

\section{Solutions in the superstring}\label{sec4}
\setcounter{equation}{0}

In this section we construct solutions
for general marginal deformations in open superstring field theory
formulated by Berkovits~\cite{Berkovits:1995ab}.
The equation of motion is
\begin{equation}\label{ssEOM}
\eta_0 \, ( \, e^{-\Phi} \, Q_B \, e^{\Phi} \, ) = 0 \,,
\end{equation}
where $\Phi$ is the superstring field of ghost number zero.
The leading term in the expansion of~(\ref{ssEOM})
in $\Phi$ is given by
\begin{equation}\label{linearized}
Q_B \, \eta_0 \, \Phi + \ord{\Phi^2} = 0 \,,
\end{equation}
where we have used $\{ Q_B , \eta_0 \} = 0$.
To leading order in the deformation parameter $\lambda$,
a solution $\Phi$ associated with an exactly marginal deformation
takes the form
\begin{equation}\label{ssLEAD}
\langle \, \varphi \,, \Phi \, \rangle
= \lambda\, \langle \, f \circ \varphi (0) \, \,
c \xi e^{-\phi} \widehat{V}_{1/2}(1) \,
\rangle_{{\cal W}_1} + \ord{\lambda^2} \,,
\end{equation}
where $\widehat{V}_{1/2}$ is the superconformal primary operator
corresponding to the marginal deformation,
as introduced in section~\ref{sec3}.
The term of~(\ref{ssLEAD}) at $\ord{\lambda}$ solves
the equation of motion to linear order in $\Phi$
given in~(\ref{linearized})
because $\eta_0$ eliminates the operator $\xi$
and the remaining unintegrated vertex operator
$c e^{-\phi} \widehat{V}_{1/2}$ in the $-1$ picture
is annihilated by the BRST operator.

In~\cite{Erler:2007rh} Erler proposed to solve
the following equation:
\begin{equation}\label{bosonictrick}
e^{-\Phi} \, Q_B \, e^{\Phi} = \Psi \,,
\end{equation}
where $\Psi$ satisfies
\begin{equation}\label{Erler-conditions}
Q_B \Psi + \Psi^2 = 0 \,, \qquad
\eta_0 \, \Psi = 0 \,,
\end{equation}
and to linear order in $\lambda$ the state $\Psi$ reduces to
\begin{equation}\label{bosLEAD}
\langle \, \varphi \,, \Psi \, \rangle
= \lambda\, \langle \, f \circ \varphi (0) \, \,
Q_B \cdot [ \, c \xi e^{-\phi} \widehat{V}_{1/2}(1) \, ] \,
\rangle_{{\cal W}_1} + \ord{\lambda^2} \,.
\end{equation}
Namely, the state $\Psi$ is
a pure-gauge string field
with respect to the gauge transformation
of {\it bosonic} string field theory,
while we use the BRST operator of the superstring.
Since the left-hand side of~(\ref{bosonictrick})
also takes a pure-gauge form,
we expect a solution $\Phi$
to the equation~(\ref{bosonictrick})
of the form (\ref{ssLEAD}).
Since $\Psi$ is annihilated by $\eta_0$,
the solution of~(\ref{bosonictrick}) also solves
the equation of motion~(\ref{ssEOM}).

In~\cite{Erler:2007rh} such a pure-gauge string field $\Psi$
was constructed from the solution
of open bosonic string field theory
in~\cite{Schnabl:2007az, Kiermaier:2007ba}
by replacing the unintegrated vertex operator
$cV_1$ in the bosonic string
with $Q_B \cdot [ \, c \xi e^{-\phi} \widehat{V}_{1/2} \, ]$
in the superstring.
Then the equation~(\ref{bosonictrick}) was solved
and solutions for marginal deformations
were constructed in open superstring field theory
when operator products of the marginal operator are
regular~\cite{Erler:2007rh, Okawa:2007it}.

We can also obtain pure-gauge string fields
satisfying~(\ref{Erler-conditions})
using the construction of solutions
for bosonic string field theory in~\cite{Kiermaier:2007vu},
which covers the case where operator products
of the marginal operator are singular.
As we reviewed in section~\ref{sec2},
the solutions in~\cite{Kiermaier:2007vu}
are constructed from the operator
$[ \, e^{\lambda V(a,b)} \, ]_r$
under the assumptions listed in appendix~\ref{assumptions}.
String fields in the superstring
satisfying~(\ref{Erler-conditions})
can be constructed from
the operator $[ \, e^{\lambda V(a,b)} \, ]_r$
with $V_1=G_{-1/2}\cdot \widehat{V}_{1/2}$
as introduced in~(\ref{ssV1}) of section~\ref{sec3}
because all the assumptions listed
in appendix~\ref{assumptions} are expected to be satisfied
when the deformation corresponding to $\widehat{V}_{1/2}$
is exactly marginal and preserves superconformal invariance.
All the solutions in section~\ref{sec2} have the same leading term
in $\lambda$ given by $\lambda \, A_L^{(1)} + \ord{\lambda^2}$,
where $A_L^{(n)}$ is defined in~(\ref{A_L/R}).\footnote{
The leading terms of the bosonic solutions
$\Psi_L$, $\Psi_R$, and $\Psi$ are
$\lambda \, A_L^{(1)}$, $\lambda \, A_R^{(1)}$,
and $\lambda \,( A_L^{(1)}+A_R^{(1)})/2$, respectively.
Because $O_R^{(1)}=O_L^{(1)}$, we have $A_R^{(1)}=A_L^{(1)}$ and thus all three solutions are equivalent
to leading order.}
As $A_L^{(1)}$ is determined by the leading term of
$[ \, O_L(a) \, e^{\lambda V(a,b)} \, ]_r$
which is given in~(\ref{leadingOs}) for the superstring case,
we find
\begin{equation}
\begin{split}
\langle \, \varphi \,, A_L^{(1)} \, \rangle
& = \langle \, f \circ \varphi (0) \, \,
O_L^{(1)} (1) \, \rangle_{{\cal W}_1}
\\
& = \langle \, f \circ \varphi (0) \, \,
[ \, cV_1(1) + \eta e^\phi \, \widehat{V}_{1/2}(1) \, ] \,
\rangle_{{\cal W}_1} \\
& = \langle \, f \circ \varphi (0) \, \,
Q_B \cdot [ \, c \xi e^{-\phi}\widehat{V}_{1/2}(1) \, ] \,
\rangle_{{\cal W}_1} \,,
\end{split}
\end{equation}
where we have used~(\ref{O1=QO1hat}).
Therefore, the condition~(\ref{bosLEAD}) is satisfied.
The solutions in~\cite{Kiermaier:2007vu} are built from
the operator $[ \, e^{\lambda V(a,b)} \, ]_r$
and its BRST transformation which are both annihilated by $\eta_0$,
and thus the second condition in~(\ref{Erler-conditions})
is also satisfied.
We can thus construct superstring solutions
for marginal deformations from the bosonic
solutions of~\cite{Kiermaier:2007vu}
by solving~(\ref{bosonictrick}).

The leading term of the superstring solution $\Phi$
in~(\ref{ssLEAD}) is built from
the leading term of the operator
$[ \, \widehat{O}_L(a) \, e^{\lambda V(a,b)} \, ]_r$,
as can be seen from~(\ref{hatO-leading}).
We therefore expect that the operator
$[ \, \widehat{O}_L(a) \, e^{\lambda V(a,b)} \, ]_r$
plays an important role in the construction
of superstring solutions.
Just as $A_L$ and $A_R$ are constructed from
the operators $[ \, O_L(1) \, e^{\lambda V(1,n)} \, ]_r$
and $[  \, e^{\lambda V(1,n)}\, O_R(n) \, ]_r$ at $\ord{\lambda^n}$,
respectively, we introduce states $\widehat{A}_L$
and $\widehat{A}_R$ which are constructed from
$[ \, \widehat{O}_L(1) \, e^{\lambda V(1,n)} \, ]_r$ and
$[  \, e^{\lambda V(1,n)}\, \widehat{O}_R(n) \, ]_r$
at $\ord{\lambda^n}$. We define
\begin{equation}
\widehat{A}_L = \sum_{n=1}^\infty \, \lambda^n \,
\widehat{A}_L^{\,(n)} \,, \quad
\widehat{A}_R = \sum_{n=1}^\infty \, \lambda^n \,
\widehat{A}_R^{\,(n)}
\end{equation}
with
\begin{equation}
\begin{split}
    \langle \, \varphi \,, \widehat{A}_L^{\,(n)} \, \rangle &=
    \phantom{-}\lim_{\epsilon\to 0}\,\sum_{l=1}^n
    \langle \, f \circ \varphi (0) \, R (1-\epsilon)
    [ \, O_L^{(l)} (1) \, V^{(n-l)} (1,n) \, ]_r \,
    \rangle_{{\cal W}_n},\\
    \langle \, \varphi \,, \widehat{A}_R^{\,(n)} \, \rangle &=
    -\lim_{\epsilon\to 0}\, \sum_{r=1}^n
    \langle \, f \circ \varphi (0) \,
    [  \, V^{(n-r)} (1,n) \, O_R^{(r)} (n)\, ]_r\, R (n+\epsilon) \,
    \rangle_{{\cal W}_n}  \,.
\end{split}
\end{equation}
The states $\widehat{A}_L$ and $\widehat{A}_R$
are related by the conjugation:
\begin{equation}\label{HatALconj}
\bigl(\widehat{A}_L\bigr)^\ddagger = {}- \widehat{A}_R \,.
\end{equation}
This can be shown as follows.
The state $R$ corresponding to the operator $R(0)$
satisfies $Q_B R = \ket{0}$ and thus $R^{\, \ddagger} = R$\,,
which follows from $\ket{0}^\ddagger = \ket{0}$
and~(\ref{conjugation-with-Q_B}).
Following the argument in~\S~2.2.1 of~\cite{Kiermaier:2007vu},
the operator $R(1-\epsilon)$ on ${\cal W}_n$
in the definition of $\widehat{A}_L^{\,(n)}$
is mapped to $R(n+\epsilon)$ under the conjugation.
The relation~(\ref{HatALconj}) then follows from
$A_L^\ddagger=A_R$.
The BRST transformations of $\widehat{A}_L$ and $\widehat{A}_R$
can be derived from those of $A_L$ and $A_R$.
The BRST transformation of $A_L$ is presented in~(\ref{Q_B-A_L}), and
using~(\ref{QU}) we find
$Q_B A_R = Q_B \, ( \, Q_B U + A_L ) = Q_B A_L$. Thus we have
\begin{equation}\label{QAL}
    Q_B A_L=-A_L \, U^{-1} \, A_R \,, \qquad
    Q_B A_R=-A_L \, U^{-1} \, A_R \,.
\end{equation}
Using the identities~(\ref{noassump1}) and~(\ref{noassump2}),
the BRST transformations of $\widehat{A}_L$ and $\widehat{A}_R$
are given by
\begin{equation}\label{identity}
    Q_B \widehat{A}_L=A_L+\widehat{A}_L \, U^{-1} \, A_R \,,
    \qquad
    Q_B \widehat{A}_R=A_R-A_L \, U^{-1} \, \widehat{A}_R \,.
\end{equation}
These relations hold
when the assumptions in appendix~\ref{assumptions} are satisfied.

We now claim that $\Phi_L$ and $\Phi_R$ defined by
\begin{equation}\label{PhiLPhiR}
e^{\Phi_L}= 1 + \widehat{A}_L \, U^{-1} \,, \qquad
e^{-\Phi_R} = 1- U^{-1} \, \widehat{A}_R
\end{equation}
solve the equation~(\ref{bosonictrick})
with $\Psi$ being $\Psi_L = A_L \, U^{-1}$
and $\Psi_R = U^{-1} \, A_R$\,, respectively,
defined in section~\ref{sec2}.
Using the relations~(\ref{QU}) and~(\ref{identity}), we have
\begin{equation}
\begin{split}
Q_B e^{\Phi_L}
&= \bigl(A_L+\widehat{A}_L \,  U^{-1} \,  A_R\bigr) \,  U^{-1}
-\widehat{A}_L \,  U^{-1} \,  (A_R-A_L) \,  U^{-1}\\
&=\bigl(1+\widehat{A}_L \,  U^{-1}\bigr) \,  A_L \,  U^{-1}\\
&=e^{\Phi_L} \,  \Psi_L \,,
\end{split}
\end{equation}
and
\begin{equation}
\begin{split}
Q_B e^{-\Phi_R}
&= {}-U^{-1} \, \bigl(A_R-A_L \,  U^{-1} \,  \widehat{A}_R\bigr)
+ U^{-1} \,  (A_R-A_L) \,  U^{-1} \, \widehat{A}_R\\
&= {}-U^{-1} \,  A_R  \, \bigl(1- U^{-1} \, \widehat{A}_R\bigr)\\
&= {}-\Psi_R \,  e^{-\Phi_R}\,.
\end{split}
\end{equation}
Therefore,
\begin{equation}
e^{-\Phi_L} \, Q_B e^{\Phi_L} = \Psi_L \,, \qquad
e^{-\Phi_R} \, Q_B e^{\Phi_R}
= {}- ( Q_B e^{-\Phi_R} ) \, e^{\Phi_R} = \Psi_R \,.
\end{equation}
Since $\Psi_L$ and $\Psi_R$ are annihilated by $\eta_0$,
the states $\Phi_L$ and $\Phi_R$ solve
the equation of motion~(\ref{ssEOM}).

The reality condition on the superstring field $\Phi$ is
$\Phi^\ddagger = {}- \Phi$, or
\begin{equation}\label{superreal}
\bigl(e^\Phi\bigl)^\ddagger=e^{-\Phi} \,.
\end{equation}
The solutions $\Phi_L$ and $\Phi_R$
do not satisfy the reality condition. In fact, we find
\begin{equation}
\bigl( e^{\Phi_L} \bigr)^\ddagger = e^{-\Phi_R} \,,
\end{equation}
which follows directly from~(\ref{HatALconj})
and the definitions~(\ref{PhiLPhiR}).
However, we can generate a real solution from
$\Phi_L$ and $\Phi_R$
by generalizing the method in appendix B of~\cite{Erler:2007rh}.
We claim that $\Phi$ defined by
\begin{equation}\label{Phireal}
e^{\Phi} = \bigl( \sqrt{e^{\Phi_L}  \,  U  \,
e^{-\Phi_R}} \, \bigr)^{-1} \,
\bigl( \, e^{\Phi_L} \, \sqrt{U} \, \bigr)
\end{equation}
satisfies the reality condition
and solves the equation of motion.
The state $\Phi$ is well defined to all orders in $\lambda$
because $e^{\Phi_L}  \,  U  \,  e^{-\Phi_R}=1+\ord{\lambda}$
and $U=1+\ord{\lambda}$.
Using the relations
\begin{equation}
e^{\Phi_L}  \,  U  \, e^{-\Phi_R}
= \bigl( \, e^{\Phi_L} \, \sqrt{U} \, \bigr) \,
\bigl( \, e^{\Phi_L} \, \sqrt{U} \, \bigr)^\ddagger \,, \qquad
\bigl( \, e^{\Phi_L} \, U \, e^{-\Phi_R} \, \bigr)^\ddagger
= e^{\Phi_L} \, U \, e^{-\Phi_R} \,,
\end{equation}
we have
\begin{equation}
\begin{split}
\bigl( e^{\Phi} \bigr)^\ddagger \, e^{\Phi}
& = \bigl( \, e^{\Phi_L} \, \sqrt{U} \, \bigr)^\ddagger \,
\Bigl( \sqrt{\bigl( \, e^{\Phi_L} \, \sqrt{U} \, \bigr)
\, \bigl( \, e^{\Phi_L} \, \sqrt{U} \, \bigr)^\ddagger} \, \,
\Bigr)^{-1} \,
\Bigl( \sqrt{\bigl( \, e^{\Phi_L} \, \sqrt{U} \, \bigr) \,
\bigl( \, e^{\Phi_L} \, \sqrt{U} \, \bigr)^\ddagger} \, \,
\Bigr)^{-1} \,
\bigl( \, e^{\Phi_L} \, \sqrt{U} \, \bigr) \\
& = \bigl( \, e^{\Phi_L} \, \sqrt{U} \, \bigr)^\ddagger \,
\Bigl( \, \bigl( \, e^{\Phi_L} \, \sqrt{U} \, \bigr) \,
\bigl( \, e^{\Phi_L} \, \sqrt{U} \, \bigr)^\ddagger \, \Bigr)^{-1} \,
\bigl( \, e^{\Phi_L} \, \sqrt{U} \, \bigr) \\
& = 1 \,.
\end{split}
\end{equation}
Therefore, $(e^{\Phi})^\ddagger=e^{-\Phi}$
and the reality condition~(\ref{superreal}) is satisfied.

The state $\Phi$ defined in~(\ref{Phireal})
is related to the solution $\Phi_L$ in the following way:
\begin{equation}
e^\Phi = \state \, e^{\Phi_L} \, \Lambda \,,
\end{equation}
where $\state = ( \sqrt{e^{\Phi_L}  \,  U  \, e^{-\Phi_R}} \, )^{-1}$
and $\Lambda = \sqrt{U}$\,.
If $\state$ is annihilated by $Q_B$
and $\Lambda$ is annihilated by $\eta_0$,
the state $\Phi$ is a gauge transformation of $\Phi_L$
and thus satisfies the equation of motion.
It is obvious that the state $\sqrt{U}$ is annihilated by $\eta_0$.
The state $( \sqrt{e^{\Phi_L}  \,  U  \, e^{-\Phi_R}} \, )^{-1}$
is annihilated by $Q_B$
if $e^{\Phi_L}  \,  U  \, e^{-\Phi_R}$ is annihilated by $Q_B$.
It can be shown as follows:
\begin{equation}
\begin{split}
Q_B \, \bigl( \, e^{\Phi_L}  \,  U  \,  e^{-\Phi_R} \, \bigr)
& = e^{\Phi_L} \, \bigl( \, \Psi_L \, U + Q_B \, U - U \, \Psi_R \,
\bigr) \, e^{-\Phi_R}
\\
& = e^{\Phi_L} \, \bigl( \, A_L  + Q_B \, U - A_R \,\bigr) \, e^{-\Phi_R} \\
& = 0 \,,
\end{split}
\end{equation}
where we used~(\ref{QU}) in the last step.
This completes the construction of real solutions
in open superstring field theory for general marginal deformations
under the assumptions listed in appendix~\ref{assumptions}.
Incidentally, the solution $\Phi$ satisfies
the equation~(\ref{bosonictrick})
with the real solution $\Psi$ of~\cite{Kiermaier:2007vu}
given in~(\ref{Psireal}):
\begin{equation}\label{Psireal-Phireal}
\begin{split}
Q_B \, e^\Phi
& = \bigl(\sqrt{e^{\Phi_L}  \,  U  \,  e^{-\Phi_R}}\,\,\bigr)^{-1} \,
Q_B \bigl( \, e^{\Phi_L} \,  \sqrt{U} \, \bigr) \\
& = \bigl(\sqrt{e^{\Phi_L}  \,  U  \,  e^{-\Phi_R}}\,\,\bigr)^{-1} \,
e^{\Phi_L} \, \bigl( \, \Psi_L \,\sqrt{U} +Q_B\sqrt{U} \, \bigr) \\
& = \bigl(\sqrt{e^{\Phi_L}  \,  U  \,  e^{-\Phi_R}}\,\,\bigr)^{-1} \,
e^{\Phi_L} \sqrt{U} \,
\Bigl( \frac{1}{\sqrt{U}}\,\Psi_L \,\sqrt{U}
+\frac{1}{\sqrt{U}}\,Q_B\sqrt{U} \, \Bigr) \\
& = e^\Phi \, \Psi \,.
\end{split}
\end{equation}
Since $\Psi$ is annihilated by $\eta_0$, we have reconfirmed that $\Phi$ solves the equation of motion~(\ref{ssEOM}).

\section{Superstring field theory around the deformed background}
\label{sec5}
\setcounter{equation}{0}

It was shown in~\cite{Kiermaier:2007vu} that
the action of open bosonic string field theory
expanded around the real solution~(\ref{Psireal})
can be written in terms of deformed algebraic structures
defined by
\begin{equation}\label{algstruc}
\begin{split}
X \star Y & \equiv X \, U^{-1} \, Y \,, \\
{\cal Q} X & \equiv
Q_B X + A_L \star X -(-1)^X \, X \star A_R
= Q_B X + \Psi_L \, X -(-1)^X \, X \, \Psi_R \,, \\
\langle\langle \, X, Y \, \rangle\rangle & \equiv
\langle \, X, U^{-1} \, Y \, U^{-1} \, \rangle
\end{split}
\end{equation}
for arbitrary string fields $X$ and $Y$.
When $\lambda= 0$,
the deformed structures reduce to their undeformed counterparts
because $ U = 1 + \ord{\lambda}$
and $A_L$, $A_R$, $\Psi_L$, and $\Psi_R$ are of $\ord{\lambda}$.
The equation of motion derived from the action
in terms of the deformed structures is
\begin{equation}
{\cal Q} \, \delta \Psi + \delta \Psi \star \delta\Psi = 0 \,,
\end{equation}
where $\delta \Psi$ is related to the original string field $\Psi$
expanded around the real solution~(\ref{Psireal}),
which we now denote by $\Psi_0$, as follows:
\begin{equation}\label{bosonic-redefinition}
\Psi = \Psi_0
+ \frac{1}{\sqrt{U}} \, \delta \Psi \, \frac{1}{\sqrt{U}} \,.
\end{equation}
The deformed structures obey the expected algebraic relations
\begin{equation}
\begin{split}
{\cal Q}^2 X & = 0 \,, \\
{\cal Q}\, ( X \star Y )
& = ( {\cal Q} X ) \star Y + (-1)^X \, X \star ( {\cal Q} Y ) \,, \\
\langle\langle \, X, Y \, \rangle\rangle
& = (-1)^{XY} \, \langle\langle \, Y, X \, \rangle\rangle \,, \\
\langle\langle \, {\cal Q} X, Y \, \rangle\rangle
& = -(-1)^{X} \langle\langle \, X, {\cal Q} Y \,\rangle\rangle \,, \\
\langle\langle \, X, Y \star Z \, \rangle\rangle
& = \langle\langle \, X \star Y , Z \, \rangle\rangle \,,
\end{split}
\end{equation}
which are necessary for a consistent formulation of string field theory.
Since open superstring field theory~\cite{Berkovits:1995ab}
is formulated using the algebraic structures
of open bosonic string field theory,
the action of open superstring field theory
written in terms of the deformed structures is consistent.
It is expected to describe fluctuations around
the background corresponding to the solution~(\ref{Phireal})
in terms of a redefined string field $\delta \Phi$.
We show in this section that this is indeed the case
and derive the relation between $\delta \Phi$
and the original string field $\Phi$
analogous to the relation~(\ref{bosonic-redefinition})
between $\Psi$ and $\delta \Psi$ for the bosonic case
found in~\cite{Kiermaier:2007vu}.

To formulate superstring field theory
using the deformed algebraic structures,
we first have to introduce
an exponential operator $\e{X}$ of the deformed star algebra.
As can be seen from the definition~(\ref{algstruc}),
the state $U$ plays the role of the identity element
of the deformed star algebra:
\begin{equation}
    X \star U = U \star X = X \,.
\end{equation}
We thus define $\e{X}$ by
\begin{equation}
\e{X} = U + X + \frac{1}{2!} \, X\star X
+ \frac{1}{3!} \, X \star X \star X + \ldots
= U + \sum_{n=1}^\infty \frac{1}{n!} \,
\underbrace{\, X \star X \star \ldots \star X \,}_{n\text{ times}} .
\end{equation}
The equation of motion derived from the action
using the deformed algebraic structures is
\begin{equation}\label{defEOM}
\begin{split}
\eta_0 \, \bigl( \,\,\e{-\delta\Phi}\star{\cal Q}\,\e{\delta\Phi}\,
\bigr) = 0 \,.
\end{split}
\end{equation}
To determine
the relation between $\delta \Phi$
and the original string field $\Phi$,
let us express~(\ref{defEOM}) in terms of the undeformed star product
and the BRST operator $Q_B$.
The exponential operator $\e{X}$ can be written as
\begin{equation}\label{}
    \e{X}= \sqrt{U}\,e^{\frac{1}{\sqrt{U}}\,X\,\frac{1}{\sqrt{U}}}\,\sqrt{U}\,.
\end{equation}
We thus find
\begin{equation}\label{deformed-to-undeformed}
\begin{split}
& \frac{1}{\sqrt{U}} \, \Bigl[ \, \eta_0 \,
\bigl( \,\,\e{-\delta\Phi}\star{\cal Q}\,\e{\delta\Phi}\, \bigr) \,
\Bigr] \, \frac{1}{\sqrt{U}} \\
& = \,\eta_0 \Bigl(\,
e^{-\frac{1}{\sqrt{U}}\,\delta\phi\,\frac{1}{\sqrt{U}}} \,
\frac{1}{\sqrt{U}}\,
{\cal Q}\,\bigl[\sqrt{U}\,
e^{\frac{1}{\sqrt{U}}\,\delta\phi\,\frac{1}{\sqrt{U}}}\,
\sqrt{U}\,\bigr]\, \frac{1}{\sqrt{U}}\,\Bigr)\,\\
& =  \,\eta_0 \Bigl(\,
e^{-\frac{1}{\sqrt{U}}\,\delta\phi\,\frac{1}{\sqrt{U}}}\,
Q_B\bigl[e^{\frac{1}{\sqrt{U}}\,\delta\phi\,\frac{1}{\sqrt{U}}}
\bigr]\,\Bigr)
+\eta_0 \Bigl(\,\bigl[Q_B\sqrt{U}-\sqrt{U}\,\Psi_R\,\bigr]\,
\frac{1}{\sqrt{U}}\,\Bigr)\,\\
& \quad ~ {}+ \,\eta_0 \Bigl(\,
e^{-\frac{1}{\sqrt{U}}\,\delta\phi\,\frac{1}{\sqrt{U}}}\,\,
\bigl[\frac{1}{\sqrt{U}}\,Q_B\sqrt{U}
+\frac{1}{\sqrt{U}}\,\Psi_L\,\sqrt{U}\,\bigr]\,
e^{\frac{1}{\sqrt{U}}\,\delta\phi\,\frac{1}{\sqrt{U}}}\,\Bigr)\,.
\end{split}
\end{equation}
The second term on the right-hand side vanishes.
Note that the real solution of bosonic string field theory
in~(\ref{Psireal}) appeared in the last line
of~(\ref{deformed-to-undeformed}).
We use the relation~(\ref{bosonictrick}) applied to the real
bosonic and superstring solutions
shown in~(\ref{Psireal-Phireal})
and find
\begin{equation}
\frac{1}{\sqrt{U}}\,Q_B\sqrt{U}+\frac{1}{\sqrt{U}}\,\Psi_L\,\sqrt{U}
= e^{-\Phi_0} \, Q_B \, e^{\Phi_0} \,,
\end{equation}
where we denoted the real superstring solution~(\ref{Phireal})
by $\Phi_0$\,. The equation of motion can then be written as
\begin{equation}
\begin{split}
& \,\eta_0 \, \Bigl(\,
e^{-\frac{1}{\sqrt{U}}\,\delta\phi\,\frac{1}{\sqrt{U}}}\,
Q_B\bigl[\,
e^{\frac{1}{\sqrt{U}}\,\delta\phi\,\frac{1}{\sqrt{U}}}\,
\bigr]
+e^{-\frac{1}{\sqrt{U}}\,\delta\phi\,\frac{1}{\sqrt{U}}}\,
\,e^{-\Phi_0} \, Q_B \bigl[\,e^{\Phi_0}\,\bigr]\,
e^{\frac{1}{\sqrt{U}}\,\delta\phi\,\frac{1}{\sqrt{U}}}\,\Bigr)\\
& = \eta_0 \, \Bigl(\,
e^{-\frac{1}{\sqrt{U}}\,\delta\phi\,\frac{1}{\sqrt{U}}}\,
\,e^{-\Phi_0} \, Q_B \bigl[\,e^{\Phi_0}\,
e^{\frac{1}{\sqrt{U}}\,\delta\phi\,\frac{1}{\sqrt{U}}}\,
\bigr]\,\Bigr) = 0 \,.
\end{split}
\end{equation}
We recognize this
as the equation of motion for the original string field $\Phi$
with the following identification:
\begin{equation}\label{excitation}
    e^\Phi = e^{\Phi_0} \,
    e^{\frac{1}{\sqrt{U}}\delta\Phi\frac{1}{\sqrt{U}}}\,.
\end{equation}
This is the relation between $\delta \Phi$ and $\Phi$,
which is a natural extension of~(\ref{bosonic-redefinition})
for the bosonic case.

\section{Discussion}\label{sec6}
\setcounter{equation}{0}

\subsection{Explicit construction for a class of marginal deformations}
\label{explicit}

We followed the strategy adopted in~\cite{Kiermaier:2007vu},
and we have presented a procedure to construct solutions
for general marginal deformations
in the superstring from properly renormalized operator products
of the marginal operator $V_1$ satisfying
the set of assumptions listed in appendix~\ref{assumptions}.
In section~4 of~\cite{Kiermaier:2007vu},
such renormalized operator products
in the bosonic string
were explicitly constructed
for a class of marginal deformations satisfying
a \emph{finiteness condition}.
To state it, we define operator products
$\dcirc V_1(t_1) \, V_1(t_2) \, \ldots \, V_1(t_{n}) \dcirc$
for arbitrary $n$ with $t_i \ne t_j$ recursively as follows:
\begin{equation}\label{NO}
\begin{split}
\dcirc V_1(t_1) \dcirc &\equiv V_1(t_1) \,, \\
\dcirc V_1(t_1) \, V_1(t_2) \, \ldots \, V_1(t_n) \dcirc
&\equiv V_1(t_1) \,
\dcirc V_1(t_2) \, \ldots \, V_1(t_n) \dcirc \\
& \quad ~ {}- \sum_{i=2}^{n} \,
\langle \, V_1(t_1) \, V_1(t_i) \, \rangle \,
\dcirc V_1(t_2) \, \ldots \,
V_1(t_{i-1}) \, V_1(t_{i+1}) \, \ldots V_1(t_n) \dcirc \,
\end{split}
\end{equation}
for $n > 1$ and $t_i\neq t_j$. The finiteness condition of~\cite{Kiermaier:2007vu} then demands that
the limit
\begin{equation}\label{finiteness}
\lim_{t \to t'} \dcirc V_1(t) \, V_1(t')^n \dcirc
\end{equation}
is finite for any positive integer $n$.

In the superstring case, we furthermore define operator products
involving the operator $\widehat{V}_{1/2}$.
For arbitrary $n$ with $t_i \ne t_j$, we define
$\dcirc \widehat{V}_{1/2}(t_1) \,
V_1(t_2) \, \ldots \, V_1(t_{n}) \dcirc$
by
\begin{equation}\label{ssNO1}
\begin{split}
\dcirc \widehat{V}_{1/2}(t_1) \, V_1(t_2) \, \ldots \,
V_1(t_n) \dcirc
&\equiv
\widehat{V}_{1/2}(t_1) \, \dcirc V_1(t_2) \, \ldots \,
V_1(t_n) \dcirc \,
\end{split}
\end{equation}
and $\dcirc \widehat{V}_{1/2}(t_1) \, \widehat{V}_{1/2}(t_2)\,
 V_1(t_3) \, \ldots \, V_1(t_{n}) \dcirc$
by
\begin{equation}\label{ssNO2}
\begin{split}
\dcirc \widehat{V}_{1/2}(t_1) \, \widehat{V}_{1/2}(t_2) \,
V_1(t_3) \, \ldots \, V_1(t_n) \dcirc
&\equiv
\widehat{V}_{1/2}(t_1) \, \dcirc \widehat{V}_{1/2}(t_2) \,
V_1(t_3) \, \ldots \, V_1(t_n) \dcirc \\
& \quad ~ {}- \langle \, \widehat{V}_{1/2}(t_1) \, \widehat{V}_{1/2}(t_2) \,
\rangle \, \dcirc  \,
V_1(t_3) \, \ldots \, V_1(t_n) \dcirc \,.
\end{split}
\end{equation}
Note that the correlation function
$\langle \, \widehat{V}_{1/2}(t) \, V_1(t') \, \rangle$ vanishes
because the conformal dimensions of the operators do not match
so that it does not appear
in the definitions~(\ref{ssNO1}) and~(\ref{ssNO2}).
Then the bosonic finiteness condition~(\ref{finiteness})
can be generalized to
the following \emph{superstring finiteness conditions}.\\

\noindent\hskip1cm
{\bf The superstring finiteness conditions.}
{\it The operators}
\begin{equation}
\lim_{t \to t'} \dcirc V_1(t) \, V_1(t')^n \dcirc \,, \qquad
\lim_{t \to t'} \dcirc \widehat{V}_{1/2}(t) \, V_1(t')^n \dcirc
\end{equation}
\hskip1cm{\it are finite for any positive integer $n$
and the operator}
\begin{equation}
\lim_{t \to t'} \dcirc \widehat{V}_{1/2}(t) \,
\widehat{V}_{1/2}(t') \, V_1(t')^n \dcirc
\end{equation}
\hskip1cm{\it vanishes
for any non-negative integer $n$.}\\

We now construct explicit solutions of superstring field theory
for the class of marginal deformations
satisfying
the superstring finiteness conditions.
The operators $[ \, e^{\lambda V(a,b)} \, ]_r$,
$[ \, V_1(a) \, e^{\lambda V(a,b)} \, ]_r$,
and $[ \, e^{\lambda V(a,b)} \, V_1(b)\, ]_r$
were explicitly constructed in~\S~4.3 of~\cite{Kiermaier:2007vu}.
When the superstring finiteness conditions are satisfied, we have
\begin{equation}
Q_B \cdot [ \, e^{\lambda V(a,b)} \, ]_r
= [ \, e^{\lambda V(a,b)} \, O_R(b) \, ]_r
- [ \, O_L(a) \, e^{\lambda V(a,b)} \, ]_r
\end{equation}
with\footnote{
When the double-pole term $1/t^2$
in the operator product expansion of $V_1(t) \, V_1(0)$
is nonvanishing, we normalize $V_1(t)$ such that the coefficient
of the double-pole term is unity.
If this convention conflicts with the reality condition
on the string field,
we set $\lambda = i \, \tilde{\lambda}$
and take $\tilde{\lambda}$ to be real
when constructing the real solution.}
\begin{equation}\label{explicitOLexpVab}
\begin{split}
[ \, O_L(a) \, e^{\lambda V(a,b)} \, ]_r
= \lambda \, c(a) \, [ \, V_1(a) \, e^{\lambda V(a,b)} \, ]_r
+ \lambda \, \eta e^\phi \, \widehat{V}_{1/2}(a) \,
[ \, e^{\lambda V(a,b)} \, ]_r
- \frac{\lambda^2}{2} \, \partial c(a) \,
[ \, e^{\lambda V(a,b)} \, ]_r \,, \\
[ \, e^{\lambda V(a,b)} \, O_R(b) \, ]_r
= \lambda \, [ \, e^{\lambda V(a,b)} \, V_1(b) \, ]_r \, c(b)
+ \lambda \, [ \, e^{\lambda V(a,b)} \, ]_r \,
\eta e^\phi \, \widehat{V}_{1/2}(b) \,
+ \frac{\lambda^2}{2} \, [ \, e^{\lambda V(a,b)} \, ]_r \,
\partial c(b) \,.
\end{split}
\end{equation}
It then follows that
\begin{equation}\label{explicithatOLexpVab}
\begin{split}
[ \, \widehat{O}_L(a) \, e^{\lambda V(a,b)} \, ]_r
= \lambda \, c \xi e^{-\phi} \, \widehat{V}_{1/2}(a) \,
[ \, e^{\lambda V(a,b)} \, ]_r
+ \frac{\lambda^2}{2} \, c \partial c\xi\partial\xi e^{-2\phi}(a) \,
[ \, e^{\lambda V(a,b)} \, ]_r \,, \\
[ \, e^{\lambda V(a,b)} \, \widehat{O}_R(b) \, ]_r
= \lambda \, [ \, e^{\lambda V(a,b)} \, ]_r \,
c \xi e^{-\phi} \, \widehat{V}_{1/2}(b) \,
- \frac{\lambda^2}{2} \, [ \, e^{\lambda V(a,b)} \, ]_r \,
c \partial c\xi\partial\xi e^{-2\phi}(b) \,.
\end{split}
\end{equation}
We can explicitly construct
superstring solutions from these operators.
By generalizing the calculation in appendix B.2
of~\cite{Kiermaier:2007vu}, we can show that
\begin{equation}
Q_B \cdot [ \,O_L(a)\, e^{\lambda V(a,b)} \, ]_r
= {}- [\,O_L(a) \, e^{\lambda V(a,b)} \, O_R(b) \, ]_r
\end{equation}
with
\begin{equation}
\begin{split}
[ \, O_L(a) \, e^{\lambda V(a,b)} \,O_L(b)\, ]_r
&=\quad \lambda^2 \, c(a) \, [ \, V_1(a) \, e^{\lambda V(a,b)}\,V_1(b) \, ]_r\,c(b)
+ \lambda^2 \, \eta e^\phi \, \widehat{V}_{1/2}(a) \,
[ \, e^{\lambda V(a,b)} \,V_1(b)\, ]_r\,c(b)\\
&\quad- \frac{\lambda^3}{2} \, \partial c(a) \,
[ \, e^{\lambda V(a,b)}\,V_1(b) \, ]_r\,c(b)
+ \lambda^2 \, c(a) \, [ \, V_1(a) \, e^{\lambda V(a,b)} \, ]_r\,\eta e^\phi \, \widehat{V}_{1/2}(b)\\
&\quad+ \lambda^2 \, \eta e^\phi \, \widehat{V}_{1/2}(a) \,
[ \, e^{\lambda V(a,b)} \, ]_r\, \eta e^\phi \, \widehat{V}_{1/2}(b)
- \frac{\lambda^3}{2} \, \partial c(a) \,
[ \, e^{\lambda V(a,b)} \, ]_r\,\eta e^\phi \, \widehat{V}_{1/2}(b)  \\
&\quad+ \frac{\lambda^3}{2} \, c(a) \, [ \, V_1(a) \, e^{\lambda V(a,b)} \, ]_r\,\partial c(b)
+ \frac{\lambda^3}{2} \, \eta e^\phi \, \widehat{V}_{1/2}(a) \,
[ \, e^{\lambda V(a,b)} \, ]_r\,\partial c(b)\\
&\quad- \frac{\lambda^4}{4} \, \partial c(a) \,
[ \, e^{\lambda V(a,b)} \, ]_r\,\partial c(b)\,,
\end{split}
\end{equation}
when the superstring finiteness conditions are satisfied.
Here the operator $[ \, V_1(a) \, e^{\lambda V(a,b)}\,V_1(b) \, ]_r$
is defined as in appendix B.1 of~\cite{Kiermaier:2007vu}.
This proves the assumption~(\ref{2})
stated in appendix~\ref{assumptions}.
The remaining assumptions~(\ref{3})--(\ref{6}) can be shown
just as in~\cite{Kiermaier:2007vu} for the bosonic case.
Thus the superstring solutions
constructed from the operators
$[ \, \widehat{O}_L(a) \, e^{\lambda V(a,b)} \, ]_r$
and $[ \, e^{\lambda V(a,b)} \, \widehat{O}_R(b) \, ]_r$
given in~(\ref{explicithatOLexpVab})
using $[ \, e^{\lambda V(a,b)} \, ]_r$
defined in section~4 of~\cite{Kiermaier:2007vu}
satisfy the equation of motion.

The simplest example of a deformation
satisfying the superstring finiteness conditions is
the deformation associated with
the constant mode of the gauge field on a D-brane in flat space.
If we denote a space-like coordinate along the D-brane by $X^\mu$
and its fermionic partner by $\psi^\mu$,
the superconformal primary field $\widehat{V}_{1/2}$
associated with the marginal deformation is given by $\psi^\mu$,
and $V_1$ is
\begin{equation}\label{simple}
\begin{split}
V_1(t) = G_{-1/2}\cdot \psi^\mu(t)
= \frac{i}{\sqrt{2 \alpha'}} \, \partial_t X^\mu (t)
\end{split}
\end{equation}
as in the bosonic case.
In this example, the operator
products~(\ref{NO}), (\ref{ssNO1}), and~(\ref{ssNO2})
are those defined by the standard normal ordering
and the superstring finiteness  conditions are
satisfied.
In~\S~4.2 of~\cite{Kiermaier:2007vu},
several examples of marginal deformations
satisfying the finiteness condition in the bosonic case
were presented. It is easy to see
by generalizing the argument in~\S~4.2 of~\cite{Kiermaier:2007vu}
that the superstring finiteness conditions are satisfied
for the supersymmetric extensions of these examples, which include
deformations of flat D-branes in flat backgrounds
by constant massless modes of the gauge field
and of the scalar fields on the D-branes,
the cosine potential for a space-like coordinate,
and the hyperbolic cosine potential
for the time-like coordinate.
Therefore, we have explicitly constructed superstring solutions for these
marginal deformations.

\subsection{More specific assumptions}\label{}

A point where the boundary condition is changed behaves
as a primary field in the bosonic case
and as a superconformal primary field
in the superstring case
and is often described in terms of
a boundary-condition changing operator.
If we assume this property,
we can derive more specific forms
of the operators
$[ \, O_L(a) \, e^{\lambda V(a,b)} \, ]_r$
and $[ \, e^{\lambda V(a,b)} \, O_R(b) \, ]_r$
both in the bosonic and superstring cases.

In the bosonic case, the BRST transformation
of a primary field $V_h (t)$ of dimension $h$ is
\begin{equation}\label{bosprimary}
Q_B \cdot V_h (t)
= c \, \partial_t V_h (t) + h \, ( \partial c) V_h (t) \,.
\end{equation}
We thus expect that
\begin{equation}\label{OLORconstrained}
\begin{split}
[ \, e^{\lambda V(a,b)} \, O_R(b) \, ]_r
& = \quad\,\, c(b) \, \partial_b \, [ \, e^{\lambda V(a,b)} \, ]_r
+ h(\lambda) \, \partial c(b) \, [ \, e^{\lambda V(a,b)} \, ]_r \,,
\\
[ \, O_L(a) \, e^{\lambda V(a,b)} \, ]_r
& = {}- c(a) \, \partial_a \, [ \, e^{\lambda V(a,b)} \, ]_r
- h(\lambda) \, \partial c(a) \, [ \, e^{\lambda V(a,b)} \, ]_r \,,
\end{split}
\end{equation}
where $h(\lambda)$ is a function of $\lambda$,
which can be interpreted as the conformal dimension
of the boundary-condition changing operator.
Therefore, once the operator $[ \, e^{\lambda V(a,b)} \, ]_r$
for arbitrary $a$ and $b$
is given, the solution is determined
up to one unknown function $h(\lambda)$.
The assumption~(\ref{2}) in appendix~\ref{assumptions}
can now be derived from~(\ref{OLORconstrained}).
We have
\begin{equation}\label{}
\begin{split}
Q_B\cdot [ \, O_L(a) \, e^{\lambda V(a,b)} \, ]_r
 = &{}-\lim_{\epsilon\to0} \,Q_B\cdot\bigl( c(a-\epsilon) \, \partial_a
+h(\lambda) \, \partial c(a-\epsilon)\bigr)\, [ \, e^{\lambda V(a,b)} \, ]_r\\
 = &{}-\,\bigl( c\del c(a) \, \partial_a
+h(\lambda) \, c\partial^2 c \bigr)\, [ \, e^{\lambda V(a,b)} \, ]_r\\
  &{}+\,\bigl( c(a) \, \partial_a+h(\lambda) \, \partial c(a)\bigr)
\,\bigl( c(a) \, \partial_a+h(\lambda) \, \partial c(a)\bigr)\, [ \, e^{\lambda V(a,b)} \, ]_r\\
  &{}+\,\bigl( c(a) \, \partial_a+h(\lambda) \, \partial c(a)\bigr)
\,\bigl( c(b) \, \partial_b+h(\lambda) \, \partial c(b)\bigr)\, [ \, e^{\lambda V(a,b)} \, ]_r\,.
\end{split}
\end{equation}
The second line on the right-hand side
precisely cancels the first line. We find
\begin{equation}\label{}
\begin{split}
Q_B\cdot [ \, O_L(a) \, e^{\lambda V(a,b)} \, ]_r
 =&\,\bigl( c(a) \, \partial_a+h(\lambda) \, \partial c(a)\bigr)
\,\bigl( c(b) \, \partial_b+h(\lambda) \, \partial c(b)\bigr)\, [ \, e^{\lambda V(a,b)} \, ]_r\\
 =&\,{}-[ \,O_L(a)\, e^{\lambda V(a,b)} \, O_R(b) \, ]_r\,,
\end{split}
\end{equation}
and thus we have derived the assumption~(\ref{2}).
The assumptions~(\ref{3})--(\ref{5})
in appendix~\ref{assumptions}
with additional operator insertions can also
be derived from those without operator insertions.

If the conformal dimension
of the boundary-condition changing operator
corresponding to a deformed background is known
and the function $h(\lambda)$ is determined from
the BRST transformation of $[ \, e^{\lambda V(a,b)} \, ]_r$,
we can identify the value of $\lambda$
which describes the deformed background.
As we discussed in~\S~\ref{explicit},
renormalized operators satisfying the assumptions
of appendix~\ref{assumptions} were constructed
in~\cite{Kiermaier:2007vu}
for a specific class of marginal deformations,
and the function $h(\lambda)$ was determined
as $h(\lambda) = \lambda^2/2$
for this class of deformations.
The simplest example
in this class is
the deformation associated with
the zero mode of the gauge field~(\ref{simple}),
and in this case the boundary-condition changing operators
at the end points $a$ and $b$ are given by
${:e^{- \frac{i \lambda}{\sqrt{2 \alpha'}}  X^\mu(a)} :}$ and
${:e^{\frac{i \lambda}{\sqrt{2 \alpha'}}  X^\mu(b)} :}$,
respectively.
They are primary fields of dimension $\lambda^2/2$,
and this is consistent with the general result
$h(\lambda) = \lambda^2/2$ for the class of deformations.
The deformation by the cosine potential~\cite{Callan:1994ub,
Polchinski:1994my, Recknagel:1998ih, Sen:1999mh}
which interpolates Neumann and Dirichlet boundary conditions
is also included in the class,
and the conformal dimension of the boundary-condition
changing operator between Neumann and Dirichlet boundary conditions
is known to be $1/16$.
Thus a natural conjecture is that
a periodic array of lower-dimensional D-branes
is described by the solution presented in section~4
of~\cite{Kiermaier:2007vu}
with $\lambda = 1/( \, 2 \sqrt{2} \, )$.\footnote{
The solution, however, is not directly constructed from
$[ \, e^{\lambda V(a,b)} \, ]_r$ but from its expansion
in $\lambda$ with different values of $a$ and $b$
for different terms in the expansion.
Furthermore, the radius of convergence in $\lambda$
of this solution is not known,
so there could be possible loopholes in our argument.}

If the renormalized operator $[ \, e^{\lambda V(a,b)} \, ]_r$
obeys the relations
\begin{equation}\label{commutederiv}
\begin{split}
\partial_a \, [ \, e^{\lambda V(a,b)} \, ]_r
&= \, [\,\partial_a \, e^{\lambda V(a,b)} \, ]_r
={}-\lambda\, [\,V_1(a)\, e^{\lambda V(a,b)} \, ]_r\,,\\
\partial_b \, [ \, e^{\lambda V(a,b)} \, ]_r
&= \, [\,\partial_b \, e^{\lambda V(a,b)} \, ]_r
=\quad\,\,\lambda\, [\, e^{\lambda V(a,b)}\,V_1(b) \, ]_r\,,
\end{split}
\end{equation}
then~(\ref{OLORconstrained})
can also be expressed as
\begin{equation}\label{OLORconstrained2}
\begin{split}
[ \, e^{\lambda V(a,b)} \, O_R(b) \, ]_r
& =   \bigl[ \, e^{\lambda V(a,b)}\,
\bigl(\lambda\,cV_1(b)
+ h(\lambda) \, \partial c(b)\bigr) \, \bigr]_r\,,\\
[ \, O_L(a) \, e^{\lambda V(a,b)} \, ]_r
& =   \, [ \,\bigl(\lambda\,cV_1(a)
- h(\lambda) \, \partial c(a)\bigr)\,
e^{\lambda V(a,b)} \, ]_r\,.
\end{split}
\end{equation}
It is easy to verify that the renormalized operators
constructed in section~4 of~\cite{Kiermaier:2007vu}
satisfy~(\ref{commutederiv}).

In the superstring case, the BRST transformation
of a superconformal primary field $\widehat{V}_h (t)$
of dimension $h$ is
\begin{equation}\label{ssprimarytrafo}
Q_B \cdot \widehat{V}_h (t)
= c \, \partial_t \widehat{V}_h (t)
+ h \, ( \partial c) \widehat{V}_h (t)
+ \eta e^\phi \, G_{-1/2} \cdot \widehat{V}_h (t) \,.
\end{equation}
Since
\begin{equation}
\begin{split}
& \lim_{\epsilon \to 0} R(t-\epsilon) \,
c(t) = 0 \,, \quad
\lim_{\epsilon \to 0} R(t-\epsilon) \, \partial c(t)
= {}- c \partial c\xi \partial \xi e^{-2 \phi} (t) \,, \quad
\lim_{\epsilon \to 0} R(t-\epsilon) \,
\eta e^\phi (t) = c \xi e^{-\phi} (t) \,,
\end{split}
\end{equation}
the ghost sectors of the operators
$[ \, \widehat{O}_L(a) \, e^{\lambda V(a,b)} \, ]_r$
and $[ \, e^{\lambda V(a,b)} \, \widehat{O}_R(b) \, ]_r$
and consequently of the superstring solutions
are highly constrained
and written in terms of $c \xi e^{-\phi}$
and $c \partial c\xi \partial \xi e^{-2 \phi}$.
We can again read off the unknown function $h(\lambda)$ from
the BRST transformation of $[ \, e^{\lambda V(a,b)} \, ]_r$
and use it to identify the value of $\lambda$
which describes a deformed background
when the conformal dimension
of the corresponding boundary-condition changing operator
is known.

\subsection{Pure-gauge forms}

As was discussed in appendix~C of~\cite{Kiermaier:2007vu},
the bosonic solutions in section~\ref{sec2}
can be formally written as pure-gauge string fields
if we use boundary-condition changing operators
expanded in $\lambda$.
Similarly, the superstring solutions in this paper
can also be formally written as pure-gauge string fields
of superstring field theory.
Let us write the operator $[ \, e^{\lambda V(a,b)} \, ]_r$ as
\begin{equation}
[ \, e^{\lambda V(a,b)} \, ]_r = \sigma_L (a) \, \sigma_R (b) \,,
\end{equation}
where $\sigma_L (a)$ and $\sigma_R (b)$
are the boundary-condition changing operators,
and expand them as follows:
\begin{equation}
\sigma_L (a) = 1+\sum_{n=1}^\infty \, \lambda^n \,
\sigma_L^{(n)} (a) \,, \qquad
\sigma_R (b) = 1+\sum_{n=1}^\infty \, \lambda^n \,
\sigma_R^{(n)} (b) \,.
\end{equation}
These are formal expansions
and we do not expect
the operators $\sigma_L^{(n)}$ and $\sigma_R^{(n)}$ to
belong to the complete set of local operators of the boundary CFT.
Then the state $U$ can be formally factorized as follows:
\begin{equation}
U = \Lambda_L \, \Lambda_R \,,
\end{equation}
where
\begin{equation}
\Lambda_L = 1 + \sum_{n=1}^\infty \lambda^n \, \Lambda_L^{(n)} \,,
\qquad
\Lambda_R = 1 + \sum_{n=1}^\infty \lambda^n \, \Lambda_R^{(n)}
\end{equation}
with
\begin{equation}
\langle \, \varphi \,, \Lambda_L^{(n)} \, \rangle
= \langle \, f \circ \varphi (0) \,
\sigma_L^{(n)} (1) \, \rangle_{{\cal W}_n} \,, \qquad
\langle \, \varphi \,, \Lambda_R^{(n)} \, \rangle
= \langle \, f \circ \varphi (0) \,
\sigma_R^{(n)} (n) \, \rangle_{{\cal W}_n} \,.
\end{equation}
The states $A_L$ and $A_R$ can be written as
\begin{equation}\label{puregaugeALAR}
A_L = {}- ( Q_B \Lambda_L ) \, \Lambda_R \,, \qquad
A_R = \Lambda_L \, ( Q_B \Lambda_R ) \,.
\end{equation}
Let us introduce states $\state_L$
and $\state_R$ defined by
\begin{equation}
\state_L
= 1 + \sum_{n=1}^\infty \lambda^n \, \state_L^{(n)} \,,
\qquad
\state_R
= 1 + \sum_{n=1}^\infty \lambda^n \, \state_R^{(n)}
\end{equation}
with
\begin{equation}
\langle \, \varphi \,, \state_L^{(n)} \, \rangle
= \langle \, f \circ \varphi (0) \, \,
Q_B \cdot [ \, R \, \sigma_L^{(n)} (1) \, ] \,
\rangle_{{\cal W}_n} \,, \qquad
\langle \, \varphi \,, \state_R^{(n)} \, \rangle
= \langle \, f \circ \varphi (0) \, \,
Q_B \cdot [ \, R \, \sigma_R^{(n)} (n) \, ] \,
\rangle_{{\cal W}_n} \,.
\end{equation}
They are obviously annihilated by the BRST operator:
$Q_B \state_L = 0$\,,
$Q_B \state_R = 0$\,.
Since
\begin{equation}
\begin{split}
\lim_{\epsilon \to 0} \,
R (a-\epsilon) \,
Q_B \cdot \sigma_L^{(n)} (a)
& = {}- \lim_{\epsilon \to 0} \,
Q_B \cdot \bigl[ \,
R(a-\epsilon) \, \sigma_L^{(n)} (a) \,
\bigr] + \sigma_L^{(n)} (a) \\
& ={}-  Q_B \cdot \bigl[ \,
R\, \sigma_L^{(n)} (a) \,
\bigr] + \sigma_L^{(n)} (a) \,,
\end{split}
\end{equation}
the state $\widehat{A}_L$ can be written as
\begin{equation}
\widehat{A}_L = \state_L \, \Lambda_R - U \,.
\end{equation}
Similarly, we have
\begin{equation}
\widehat{A}_R = U - \Lambda_L \, \state_R  \,.
\end{equation}
Thus the solutions $e^{\Phi_L}$ and $e^{-\Phi_R}$ can be written as
\begin{equation}
\begin{split}
e^{\Phi_L}
& = 1 + ( \, \state_L \, \Lambda_R - U \, ) \, U^{-1}
= \state_L \, \Lambda_L^{-1} \,, \\
e^{-\Phi_R}
& = 1 - U^{-1} \,
( \, U - \Lambda_L \, \state_R \, )
= \Lambda_R^{-1} \, \state_R \,.
\end{split}
\end{equation}
The left factor $\state_L$ of $e^{\Phi_L}$
and the right factor $\state_R$ of $e^{-\Phi_R}$
are annihilated by $Q_B$.
Furthermore, the right factor $\Lambda_L^{-1}$ of $e^{\Phi_L}$
and the left factor $\Lambda_R^{-1}$ of $e^{-\Phi_R}$
are annihilated by $\eta_0$
so that both $\Phi_L$ and $\Phi_R$ are formally written
as pure-gauge string fields of superstring field theory.
For the particular marginal deformation~(\ref{simple})
associated with turning on the zero mode of the gauge field,
$\Phi_L$ corresponds to the solution
constructed in~\cite{Fuchs:2007gw}.

We can also express the real solution $\Phi$ in~(\ref{Phireal})
in terms of $\Lambda_L$, $\Lambda_R$, $\state_L$, and $\state_R$.
Since
\begin{equation}
e^{\Phi_L}\,U\,e^{-\Phi_R}
= \state_L \, \state_R \,,
\end{equation}
we have
\begin{equation}
e^{\Phi}
= \bigl( \sqrt{e^{\Phi_L}  \,  U  \,
e^{-\Phi_R}}\,\,\bigr)^{-1} \,
\bigl(e^{\Phi_L} \, \sqrt{U} \, \bigr)
= \Bigl[ \, \bigl( \, \sqrt{ \, \state_L \,
\state_R} \, \bigr)^{-1} \,
\state_L \, \Bigr] \,
\Bigl[ \, \Lambda_L^{-1} \, \sqrt{\Lambda_L\,\Lambda_R} \,
\Bigr] \,.
\end{equation}
This expression for $e^\Phi$ is again formally
in a pure-gauge form
because the left factor is annihilated by $Q_B$
and the right factor is annihilated by $\eta_0$.
Thus we have also solved the problem of finding
a real superstring solution in a pure-gauge form
raised in~\cite{Fuchs:2007gw}.

\vskip 1cm

\noindent
{\bf \large Acknowledgments}

\medskip

We would like to thank
Volker Schomerus and Barton Zwiebach
for useful discussions.
The work of M.K. is supported in part
by the U.S. DOE grant DE-FG02-05ER41360
and by an MIT Presidential Fellowship.

\bigskip

\appendix

\section{Assumptions}
\label{assumptions}
\setcounter{equation}{0}

In this appendix we present a list of the assumptions
introduced in~\cite{Kiermaier:2007vu}
on the renormalized operator
$[ \, e^{\lambda V(a,b)} \, ]_r$
for constructing solutions
corresponding to general marginal deformations.
See~\S~1.1 of~\cite{Kiermaier:2007vu}
for more detailed discussion.
While the discussion in~\cite{Kiermaier:2007vu}
was for the bosonic string,
it can be extended to the superstring
if the marginal operator $V_1$
is the supersymmetry transformation
of a superconformal primary field $\widehat{V}_{1/2}$
of dimension $1/2$
as stated in~(\ref{ssV1})
and if the BRST operator~(\ref{Q_B})
for the superstring is used.
We believe that all the assumptions
are satisfied for any exactly marginal deformation
preserving superconformal invariance.\\

\noindent
{\it 1. The BRST transformation of
the operator $[ \, e^{\lambda V(a,b)} \, ]_r$
takes the following form:}
\begin{equation}
Q_B \cdot [ \, e^{\lambda V(a,b)} \, ]_r
= [ \, e^{\lambda V(a,b)} \, O_R (b) \, ]_r
- [ \, O_L (a) \, e^{\lambda V(a,b)} \, ]_r \,,
\tag{I}\label{1}
\end{equation}
{\it where $O_L (a)$ and $O_R (b)$ are some local operators
at $a$ and $b$, respectively.}\\

\noindent
{\it 2. The BRST transformation of
the operator $[ \, O_L (a) \, e^{\lambda V(a,b)} \, ]_r$
is given by}
\begin{equation}
Q_B \cdot [ \, O_L (a) \, e^{\lambda V(a,b)} \, ]_r
= {}- [ \, O_L (a) \, e^{\lambda V(a,b)} \, O_R (b) \, ]_r \,.
\tag{II}\label{2}
\end{equation}\\
The operator $[ \, e^{\lambda V(a,b)} \, ]_r$
generalizes to
\begin{equation}
[ \, \prod_{i=1}^n \, e^{\lambda_i V (a_i, a_{i+1} )} \, ]_r \,
\label{multiple-regions}
\end{equation}
with $a_i < a_{i+1}$ for $i = 1, 2, \ldots , n$
when different boundary conditions
on different segments on the boundary are introduced.
Two assumptions on this operator were made
in~\cite{Kiermaier:2007vu}.\\

\noindent
{\it 3. Replacement. When $\lambda_{i+1} = \lambda_i$, the product
$e^{\lambda_i V (a_i, a_{i+1} )} \,
e^{\lambda_{i+1} V (a_{i+1}, a_{i+2} )}$
inside the operator (\ref{multiple-regions})
can be replaced by
$e^{\lambda_i V (a_i, a_{i+2} )}$}:
\begin{equation}
[ \, \ldots \, e^{\lambda_i V (a_i, a_{i+1} )} \,
e^{\lambda_i V (a_{i+1}, a_{i+2} )} \, \ldots \, ]_r
= [ \, \ldots \, e^{\lambda_i V (a_i, a_{i+2} )} \,
\ldots \, ]_r \,.
\tag{III}\label{3}
\end{equation}\\
\noindent
{\it 4. Factorization. When $\lambda_j$ vanishes,
the renormalized product~(\ref{multiple-regions})
factorizes as follows:}
\begin{equation}
[ \, \ldots \, e^{\lambda_{j-1} V (a_{j-1}, a_{j} )} \,
e^{\lambda_{j+1} V (a_{j+1}, a_{j+2} )} \, \ldots \, ]_r
=
[ \, \ldots \, e^{\lambda_{j-1} V (a_{j-1}, a_{j} )} \,]_r\,
[ \, e^{\lambda_{j+1} V (a_{j+1}, a_{j+2} )} \, \ldots \, ]_r
\,.
\tag{IV}\label{4}
\end{equation}\\
\noindent
It was also assumed that (\ref{3}) and (\ref{4}) hold
when $O_L (a_1)$, $O_R (a_{n+1})$ or both are inserted
into the operator~(\ref{multiple-regions}).
The next assumption is for operators
on the family of surfaces ${\cal W}_n$.\\

\noindent
{\it 5. Locality. The operators $[ \, e^{\lambda V(a,b)} \, ]_r$ and
 $[ \, O_L (a) \, e^{\lambda V(a,b)} \, ]_r$
defined on ${\cal W}_n$ coincide
with those defined on ${\cal W}_m$ with $m > n$}:
\begin{equation}
\begin{split}
[ \, e^{\lambda V(a,b)} \, ]_r
~ \text{on} ~ {\cal W}_n ~
& = ~ [ \, e^{\lambda V(a,b)} \, ]_r
~ \text{on} ~ {\cal W}_m \,, \\
[ \, O_L(a) \, e^{\lambda V(a,b)} \, ]_r
~ \text{on} ~ {\cal W}_n ~
& = ~ [ \, O_L(a) \, e^{\lambda V(a,b)} \, ]_r
~ \text{on} ~ {\cal W}_m \,.
\end{split}
\tag{V}\label{5}
\end{equation}\\
Finally, $e^{\lambda V(a,b)}$
is classically invariant under the reflection
where $V_1(t)$ is replaced by $V_1(a+b-t)$,
and it was assumed that
$[ \, e^{\lambda V(a,b)} \, ]_r$ preserves this symmetry.\\

\noindent
{\it 6. Reflection. The operator $[ \, e^{\lambda V(a,b)} \, ]_r$
is invariant under the reflection
where $V_1(t)$ is replaced by $V_1(a+b-t)$}:
\begin{equation}
\biggl[ \, \exp \biggl( \, \lambda \int_a^b dt \, V_1(a+b-t) \,
\biggr) \, \biggr]_r
= \biggl[ \, \exp \biggl( \, \lambda \int_a^b dt \, V_1(t) \,
\biggr) \, \biggr]_r \,.
\tag{VI}\label{6}
\end{equation}\\

\end{document}